\documentstyle[12pt,epsf]{article}
\newlength{\abstwidth}
\def\nn{\nonumber}
\def\als{\alpha_s}

\begin{document}
\pagestyle{empty}

\begin{flushright}
hep-ph/9807529\\
JINR E2-98-223\\
WU B 98-24\\
July 1998
\end{flushright}

\vspace{\fill}
\begin{center}
{\large{\bf Spin effects in high-energy proton-proton scattering within
a diquark model}}\\[10mm]
{\Large S.V.Goloskokov} \footnote{Email:
goloskkv@thsun1.jinr.dubna.su}\\ {\it Bogoliubov Laboratory of
Theoretical
  Physics, Joint Institute for\\  Nuclear Research, Dubna
  141980, Moscow
  region, Russia.}\\[1mm]
{\Large P. Kroll}\footnote
{kroll@theorie.physik.uni-wuppertal.de}\\
{\it Fachbereich Physik, Universit\"at Wuppertal, \\
D-42097 Wuppertal, Germany}
\end{center}
\vspace{\fill}
\begin{center}
{\bf Abstract}\\[2ex]
\begin{minipage}{\abstwidth}
We study $pp$ scattering at high energies and moderately large
momentum transfer, using a model in which the proton is viewed as being
composed of a quark and a diquark. We show that this model leads to
single and double spin transverse asymmetries which are neither small
nor vanish at high energies.
\end{minipage}
\end{center}
\vspace{\fill}
\newpage
\pagestyle{plain}
\setcounter{page}{1}
\section{Introduction}
The adequate theoretical description of spin effects in high-energy exclusive
processes at moderately large momentum transfer is one of the unsolved
problems in QCD. As is well known, massless QCD,
leads to hadronic helicity conservation and, hence, to zero
single-spin asymmetries. Mass  and higher order perturbative QCD
corrections lead to a non-vanishing single-spin transverse asymmetries:
\begin{equation}
\label{anqcd}
 A_N \ \propto  m \alpha_s / \sqrt{-t}.
\end{equation}
A QCD analysis reveals that the mass parameter $m$ appearing in
(\ref{anqcd}) is of order of the hadron mass \cite{ter} and should not be
interpreted as a current quark mass. So, one may expect a substantial
single-spin asymmetry for momentum transfer, $-t$, of the order of
a few GeV$^2$. Actual estimates within QCD inspired models
provide only values of the order of a few per cents for
single-spin asymmetries, indeed much smaller than the experimental results.

Experimentally, there are many observations of large spin effects
at high energies and moderately large momentum transfer \cite{hesp}.
Sizeable differences between the cross sections for different spin
orientations of the initial state protons as well as large
double-spin, $A_{NN}$, and single-spin, $A_N$, transverse asymmetries have
been observed in the BNL experiment \cite{krish} for
beam momenta $p_B$ less than $28\;{\rm GeV}$. The FNAL experiment
\cite{fnalp} finds values for $A_N$ of about 10-20\% at $p_B=200\;{\rm
GeV}$ and momentum transfers $|t| \ge 2\;{\rm GeV}^2$.
This result is of the same order of magnitude than the BNL asymmetry at
$p_B= 28\;{\rm GeV}$  and similar values of $t$.
Combining these observations with corresponding ones made at
small momentum transfer \cite{akch}, one is lead to the conclusion that
spin effects in high-energy reactions exhibit a weak energy
dependence.

Elastic scattering at high energies and fixed momentum transfer
($|t|/s$ small) is customarily believed to be under control of
the $t$-channel colour-singlet Pomeron (and, eventually Odderon)
exchange that has a dominant non-flip coupling. The observed spin
effects thus seem to require the existence of an additional
Pomeron-like exchange in the helicity-flip amplitudes that has -
up to eventual $\ln{s}$ factors - the same energy dependence as
the standard Pomeron but is not in phase with it. Within QCD the
Pomeron is interpreted as $t$ channel exchange of gluons with
total charge conjugation of unity ($C=+1$). Present attempts to
understand it theoretically are based on the simple two-gluon
picture for this object \cite{low}. It is important to note that
in such models the Pomeron couples to quarks and not directly to
the hadrons. According to the model \cite{lansh-m}, the gluons
representing the Pomeron preferentially interact with the same
quark within a given hadron. As a consequence of this property,
the Pomeron effectively couples to the hadron like an $C= +1$
isoscalar photon \cite{lansh-m} and approximately reproduces the
salient features of the additive quark model. In the
Landshoff-Nachtmann (LN) non-perturbative model \cite{la-na},
the two gluon representing the Pomeron do not only couple to one
and the same constituent. However, neither the LN model
\cite{la-na}  nor that of \cite{lansh-m} provides a
spin-dependent Pomeron coupling. The question of gauge invariance
for the models \cite{lansh-m,la-na} has been investigated by
Diehl \cite{diehl}.

In several models high energy spin effects have been investigated.
Thus, for instance, in \cite{gol-93} the spin-dependent quark-Pomeron
coupling was constructed from a gluon-loop contribution. It was shown
that this quark-Pomeron coupling leads to fairly large spin
asymmetries in diffractive quark-antiquark pair
production and exhibits only a weak energy dependence \cite{gol-96}.
In \cite{models} rotating matter inside the proton was claimed to be the
origin of spin effects. The authors of \cite{gol-91} considered the
Pomeron interaction with the  light quark-antiquark cloud of the
proton. While these models provide spin effects at high
energies in fair agreement with experiment they suffer from the large
number of adjustable parameters they depend on. Moreover, the
applicability of these models is restricted to small momentum transfer.

Here, in this work, we are interested in spin effects at high energies
and moderately large momentum transfer ($3\;{\rm GeV}^2 <|t| <<s$).
In view of the polarization physics programs proposed for the future
proton accelerators \cite{prop} this kinematical region is of
topical interest. Our approach is based on the diquark picture
\cite{kroll} where the proton is viewed as being composed of a quark
and a diquark in the dominant valence Fock state instead of three
quarks. The diquarks represent an effective description of
non-perturbative effects; their composite nature is taken into account
by diquark form factors. The diquark picture of the proton simplifies
our calculations drastically due to the reduced number of
constituents. The combination of the quark-diquark picture of the
proton and the hard scattering approach developed by Brodsky and
Lepage \cite{Bro:80} leads to successful descriptions of
electromagnetic form factors and other exclusive reactions
\cite{diquarks,Jak:93b} at fairly large momentum transfer.
Spin effects are generated from spin 1 (vector) diquarks in that
model. The model also provides phase differences between different helicity
amplitudes in some cases and can therefore account for single-spin
asymmetries in principle.
Note, that these corrections are non-Pomeron like because of
the phase shift between the flip and non-flip contributions.
However, even within the diquark model which is much simpler to handle than
the three-quark picture of the proton, a full hard scattering
analysis of elastic proton-proton scattering is beyond
feasibility at present (see, for instance, \cite{jacob}). Therefore,
in order to simplify and in regard to the fact that we are not
interested in the real hard scattering region for which the diquark
model was originally designed, we use that model in combination with the
two-gluon exchange picture as a representative of the Pomeron. We
calculate the helicity-flip amplitude explicitly in that framework while,
at the end, the non-flip amplitudes are described by a standard
phenomenological Pomeron exchange.
We note that Ramsey and Sivers \cite{ramsey} also proposed a hard
scattering model that produces substantial spin effects. This model is
based on quark-exchange and the Landshoff pinch contribution
\cite{landsh-p} to the $pp$ helicity amplitudes.

In Sect.\ 2 we begin with a few kinematical preliminaries. A brief
description of the diquark model is presented in Sect.\ 3. The general
structure of the various diquark contributions to elastic $pp$
scattering is discussed in Sect.\ 4. In Sect.\ 5 we present our
numerical results for spin asymmetries in elastic $pp$ scattering
and compare them to experimental data. Concluding remarks are given in
Sect.\ 6.

%
\section{Proton-proton scattering at high energies}
The momenta and the Mandelstam variables of elastic proton--proton
scattering are defined by
\begin{equation}
p(p_1)+p(p_2) \to p(p_3)+p(p_4)  \label{pp}
\end{equation}
and
$$ s=(p_1+p_2)^2, \quad  t=(p_1-p_3)^2. $$

Elastic $pp$ scattering can be described in terms  of helicity
amplitudes
\begin{equation}
T_{\lambda_4 \lambda_3;\lambda_2 \lambda_1} =
\bar u(p_4,\lambda_4) \bar u(p_3,\lambda_3)\hat T(s,t)
u(p_2,\lambda_2) u(p_1,\lambda_1).
\label{hel}
\end{equation}
of which only five are independent. In (\ref{hel}) $u$ denotes
the spinor of a proton with momentum $p_i$ and helicity
$\lambda_i$. In the kinematical region of interest the double
helicity--flip amplitudes are believed to be much smaller than
the helicity non--flip ones and the two non-flip amplitudes are
of equal magnitude approximately. These properties hold in most
of models (see, for instance, \cite{models,gol-91}) and we will
assume that they also hold in our approach. In this situation we
can, for convenience and without loss of generality, fix the
helicities of the protons 1 and 3 at $+1/2$. Therefore, we have
 to model a non-flip, $F_{++}$, and a flip amplitude, $F_{+-}$,
only. $F_{++}$ represents the average of the two non-flip
amplitudes. There is no need for antisymmetrization of the
amplitudes since the $p_3 \leftrightarrow p_4$ interchanged
contribution is suppressed by inverse powers of $s$ in the
kinematical region of interest ($t\leftrightarrow u \simeq s$).

In terms of the amplitudes $F_{++}$ and $F_{+-}$ the differential cross
sections is given by
\begin{equation}
\frac{d \sigma}{dt}=\frac{1}{64 \pi s^2}
[|F_{++}|^2+2 |F_{+-}|^2].
\end{equation}

The single-spin asymmetry reads
\begin{equation}
\label{an}
A_N= -2 \frac{{\rm Im}[F_{++} F_{+-}^*]}{|F_{++}|^2+2 |F_{+-}|^2}
\end{equation}
while  the double spin transverse asymmetry is given by
\begin{equation}
\label{ann}
A_{NN}=  2  \frac{|F_{+-}|^2}{|F_{++}|^2+2 |F_{+-}|^2}.
\end{equation}

The $A_{NN}$ asymmetry is related to the differential cross-sections
in parallel and anti-parallel spin states by
\begin{equation}
\label{dsparapar}
\frac{d\sigma(\uparrow\uparrow)/dt}{d\sigma(\uparrow
\downarrow)/dt}=\frac{1+A_{NN}}{1-A_{NN}}.
\end{equation}

In the following we are going to calculate the leading
contribution to   the helicity-flip amplitude within the
diquark model, omitting corrections like $m^2/t$. The
non-flip amplitude, on the other hand, is modelled by a
phenomenological ansatz. As a crossing-even exchange the grows
$\propto s$, the Pomeron contribution is dominantly imaginary
with only a very small real part suppressed by $1/s$ as follows
from analyticity \cite{coll}. We will make use of two
alternative parametrizations valid for $|t|$ larger then
$3\;{\rm GeV}^2$ (after the dip region of the differential cross
section): Following, for instance, the authors of
\cite{gol-91}, we parametrize $F_{++}$ as an exponential
\begin{equation}
\label{mpe}
F_{++}(s,t) = i s\  b \exp{(-a \sqrt{|t|})}.
\end{equation}
This ansatz is understood as being a consequence of multiple
Pomeron exchange (MPE). Alternatively, we use
the parametrization
\begin{equation}
\label{pinch}
F_{++}(s,t) = i s \frac{c}{t^4}.
\end{equation}
which may be viewed as a phenomenological version of the
Landshoff pinch contribution (LP) \cite{landsh-p} to $pp$
scattering. Note, that the model results
\cite{gol-91,landsh-p} confirm the imaginary of the amplitudes
(\ref{mpe},\ref{pinch}). In our numerical estimations we shall
use the MPE fit for $b= -45.967 \;{\rm GeV^{-2}}, a= 3.745 \;{\rm
GeV^{-1}}$ and the LP fit for $c=  -6.284 \;{\rm GeV^{6}} $. Both
the parametrizations, (\ref{mpe}) and (\ref{pinch}), describe
rather well the $pp$ differential cross section data at ISR
energies \cite{spps}. An eventual residual energy dependence of
the experimental data (perhaps of $\ln{s}$ type) will be ignored
here. It is irrelevant for our purpose of investigating spin
effects.
%
\section{The diquark model}
As we said in the introduction we will make use of the diquark
model of the proton advocated for in \cite{kroll,diquarks,Jak:93b}.
Here we give a brief description of that model.
In the hard scattering approach proposed by Brodsky and Lepage \cite{Bro:80}
the process $p\,p \rightarrow p\,p$
is expressed by a convolution of distribution amplitudes (DA)
with hard-scattering amplitudes
calculated in collinear approximation within perturbative QCD.
In a collinear situation in which intrinsic transverse momenta
are neglected and all constituents of a hadron have momenta
parallel to each other and parallel to the momentum of the
parent
hadron, one may write the valence Fock state of the proton in a
covariant
fashion (omitting colour indices for convenience)
\begin{equation}
\label{pwf}
|p,\lambda:qS;qV,{\alpha}\rangle  = f_S\,\varphi_S(x_1)\,B_S\, u(p,\lambda)
             + f_V\, \varphi_V(x_1)\, B_V
              (\gamma^{\alpha}+p^{\alpha}/m)\gamma_5
              \,u(p,\lambda)/\sqrt{3}\, .
\end{equation}
The Lorentz index $\alpha$ represents
the polarization state of the vector diquark.
The two terms in (\ref{pwf}) represent configurations
consisting of a quark and either a spin-isospin zero $(S)$ or a
spin-isospin one $(V)$ diquark, respectively. The couplings of
the diquark
with the quarks in a proton lead to the flavour functions
\begin{equation}
\label{fwf}
B_S=u\, S_{[u,d]}\, ,\hspace{2cm}
B_V= [ u V_{\{u,d\}} -\sqrt{2} d\, V_{\{u,u\}}]/\sqrt{3}\, ,
\end{equation}
where the subscripts indicate the flavour content of the diquarks
($S$,$V$) in either antisymmetric or symmetric combinations.
The DA $\varphi_{S(V)}(x_1)$, where $x_1$ is the momentum
fraction carried
by the quark, represents the light-cone wave function integrated
over
transverse momentum and is defined in such way that
\begin{equation}
\label{DAn}
\int_0^1 dx_1\,\varphi_{S,(V)}(x_1)=1\;.
\end{equation}
The constant $f_{S(V)}$ acts as the value of the configuration
space wave function at the origin.

The  amplitude  $F_{+-}$  will be calculated in the spirit of the
hard scattering approach \cite{Bro:80} where the quarks and diquarks
are connected by the minimal number of gluons, i.e.\ by three.
Disconnected Feynman graphs are suppressed in the
kinematical region of interest \cite{Bro:80}. We
also will employ several kinematical simplifications since we only
consider the region $m^2<<|t|<<s$. Colour neutralization requires the
$t$-channel exchange of two gluons. The third one is exchanged within
one of the proton-proton vertices. In so far our model for the flip
amplitude bears resemblance to the Landshoff-Nachtmann \cite{la-na}
two-gluon model of the Pomeron.
In contrast to \cite{lansh-m}
which refers to the standard non-flip Pomeron at small -t,
in our approach the two gluons exchanged between the two proton-proton
vertices do not only couple to one and the same constituent. This is
not a contradiction since we are interested in a helicity-flip
amplitude at high energies and moderately large momentum transfer.
The helicity-flip amplitude can be
expressed as a product of a helicity non-flip vertex (HNF) and flip
vertex (HF). The structure of the HNF vertex is shown in Fig.\ 1. For
this vertex we only consider scalar diquarks in order to keep the
model simple. The graphs contributing to the product of the HNF and
the HF are shown in Figs.\ 2--5. To the HF vertex only vector
diquarks contribute since, obviously, from scalar diquarks a helicity
flip cannot be generated. The graphs shown in Figs.\ 2 and 3 contain
3-point diquark vertex functions while those shown in Figs. 4
(three-gluon interactions) and 5 (without three-gluon interactions)
contain 4-point functions. In principle there is also a graph with a
quartic gluon coupling. However, its contribution is suppressed at
large $s$. It has been shown in \cite{jacob} that this set
of graphs leads to gauge-invariant scattering amplitudes.
The n-point functions, indicated by blobs in Figs.\ 2--5, are given by
a product of the relevant graphs for point-like diquarks (see, for instance,
Fig.\ 6) and appropriate phenomenological diquark form factors. These
form factors take into account the composite nature of the diquarks.
Since the 5-point functions provide only small corrections to the
final results we omit them in our analysis.

The perturbative part of the diquark model, i.e.\ the coupling
of gluons to diquarks follows standard
prescriptions (for notations refer to \cite{Jak:93b})
\begin{eqnarray}
&& \mbox{SgS}: i\,g_s t_{ij}^{a}\,(p_1+p_2)_{\mu} \nonumber\\
&& \mbox{VgV}: -i\,g_{s}t_{ij}^{a}\,
\left\{
 g_{\alpha\beta}(p_1+p_2)_{\mu}
- g_{\mu\alpha}\left[(1+\kappa)\,p_1-\kappa\, p_2\right]_{\beta}
- g_{\mu\beta} \left[(1+\kappa)\,p_2-\kappa\,
p_1\right]_{\alpha}
\right\}
\end{eqnarray}
where $g_s=\sqrt{4\pi\alpha_s}$ is the QCD coupling constant.
$\kappa$ is the anomalous magnetic moment of the vector diquark
and $t^a=\lambda^a/2$ the Gell-Mann colour matrix.  The couplings
$DgD$ are supplemented by appropriate contact terms required by
gauge invariance, e.g.\
\begin{equation}
 \mbox{gSgS}: -i\, g_s^2 \{t^a_,t^b\}_{ij}\, g_{\mu\nu}
\end{equation}

The phenomenological diquark form factors are taken from
\cite{kroll,diquarks}
\begin{equation}
\label{fs3}
F_{S}^{(3)}(Q^{2})=\frac{Q_{S}^{2}}{Q_{S}^{2}+Q^{2}};\qquad
F_{V}^{(3)}(Q^{2})=\left(\frac{Q_{V}^{2}}{Q_{V}^{2}+Q^{2}}\right
)^{2}\, ;
\end{equation}
\begin{equation}
\label{fsn}
F_{S}^{(4)}(Q^{2})=a_{S}F_{S}^{(3)}(Q^{2}); \qquad
F_{V}^{(4)}(Q^{2})=a_{V}\left(\frac{Q_{V}^{2}}{Q_{V}^{2}+Q^{2}}\right
)^{3}\,.
\end{equation}
The constants $a_{S}$ and $a_{V}$ are strength parameters introduced
in order to take care of diquark excitation and break-up. These
parametrizations are constrained by the requirement that
asymptotically the diquark models evolves into the standard
Brodsky-Lepage hard scattering model \cite{Bro:80}.
%
\section{The structure of the model amplitude}
According to our discussion in Sect.\ 3 the helicity flip amplitude can
be expressed as a product of the helicity non-flip vertex to which
only scalar diquarks contribute and the flip vertex that, in our
model, is controlled by vector diquarks:
\begin{eqnarray}
F_{+-}(s,t)&=& s\, \sqrt{-t}\,\frac{(4\pi)^3}{3t^2}\,f_S^2\,f_V^2 \nn\\
      &\times & \int d\alpha_1 d\beta_1 \frac{\phi_S(\alpha_1)
                \phi_S(\beta_1)}{\alpha_1\alpha_2\beta_1\beta_2}
                \, \als(-\alpha_1\beta_1 t) \als(-\alpha_2\beta_2 t)
                \, F_S^{(3)}(-\alpha_2\beta_2 t) \nn\\
      &\times &\int dx_1 dy_1 \phi_V(x_1) \phi_V(y_1) \sum_i C_i \hat A_i
\label{ampl}
\end{eqnarray}
$\alpha_1$ and $\beta_1$ denote the fractions of the baryon
momentum carried by the quarks in the initial and final baryons
entering the HNF-vertex, respectively. $\alpha_2=1-\alpha_1$ and
$\beta_2=1-\beta_1$ are the momentum fractions the diquarks carry.
$x_1, (x_2), y_1 (y_2)$ are the analogue quantities for the
HF-vertex. $C_i$ is the color factor. To facilitate the discussion
we split $F_{+-}$ into
contributions from various groups of Feynman graphs. The $\hat A_i$
are written as a contraction of the two tensors representing the
HNF and HF vertices
\begin{equation}
\hat A_i \,=\, H^{n.f.}_{\mu\nu}\;\cdot H^{\mu\nu}_{fi}
\end{equation}
The HNF tensor has the simple form
\begin{equation}
 H^{n.f.}_{\mu\nu} = \bar u(p_3+) [\gamma_{\nu} (p_1+p_3)_{\mu} \,+\,
                          \gamma_{\mu} (p_1+p_3)_{\nu} ] u(p_1,+)
\end{equation}
The HF tensors are to be calculated from the Feynman graphs shown in
Figs.\ 2-6. They contain a factor of $\als$
with an appropriate argument (representing the virtuality of the
internal gluon) and the vector diquark form factor besides the
characteristics of the relevant Feynman graphs. We refrain from
quoting the $H_{fi}^{\mu}$ explicitly but discuss the functions the
functions $\hat A_i$ directly.

The graph 2a includes a propagator (marked by a cross) whose denominator
contains a term proportional to $s$. Neglecting  in this denominator
terms proportional to $t$ and $m^2$ in accordance with the condition
$m^2, |t| \ll s$, we have
\begin{eqnarray}
\hat A_{(2a)}& =&\hat a_{(2a)}(\alpha_1,\beta_1)
           \left[\frac{1}{s y_1 (\alpha_1-\beta_1) +{\rm i}\,\epsilon}
            +\frac{1}{-s y_1 (\alpha_1-\beta_1) +{\rm i}\, \epsilon}\right]
\nonumber \\[0.5ex]
  &=& -\frac{2 {\rm i} \pi}{s y_1}\,\hat a_{(2a)}(\alpha_1,\alpha_1)\,
\delta(\alpha_1-\beta_1).
\end{eqnarray}
where the regular function $\hat a_{2a}(\alpha_1,\alpha_1)$ is given in Tab.\
\ref{tab}. The contribution from graph 2b is given by $\hat
A_{(2a)}(x_1,y_1)=\hat A_{(2b)}(y_1,x_1)$.
\renewcommand{\arraystretch}{1.0}
\begin{table}
\begin{tabular}{|c|c|p{12.5cm}|}\hline
 Graph&$C_i$& \\ \hline
$  2a $&$ \frac{8}{27} $
&$\hat a_{(2a)}=-\frac{2\,s^2\,y_1\,\alpha_s(-x_2 y_2 t) \alpha_2 \left[
2(x_2+y_2) -\kappa(3  x_1- 2 y_2 )\right]}{t m x_2 y_2 }  F_V^{(3)}(-x_2
y_2 t)
$
\\[1.0ex] \hline
$  3a $&$\frac{i}{3}$&$
\hat a_{(3a)}= \frac{-2 \, s^2\, \alpha_s(-x_2 y_2 t)\, \alpha_2 }{ m x_2 y_2}
[2 (x_2+y_2) (2 y_1^2-y_1 \alpha_2-2 x_1 y_1+ 2 x_1 \alpha_2 +\alpha_1^2-1)$
                                                          \\[1.0ex]
&&$- \kappa (5 x_1^2+4 y_1^3-10 y_1^2+10 y_1-5 x_1^2 y_1+3 x_1
   y_1^2-4- 3 x_1 y_1+4 \alpha_1^2 $\\
&&$-5 x_1^2 \alpha_1 -2 x_1 \alpha_1^2+2 x_1 \alpha_1+6y_1^2 \alpha_1-
2 y_1 \alpha_1^2-4 y_1 \alpha_1+x_1 y_1 \alpha_1 ] F_V^{(3)}(-x_2 y_2
t)$
\\ [1.0ex] \cline{3-3}
&&
\footnotesize{$d_{(3a)1}=(\alpha_1-x_2) (\alpha_1-y_2) t+(x_2-y_2)^2 m^2$}\\
&&
\footnotesize{$d_{(3a)2}=-(\alpha_1-y_2) \alpha_2 t+y_1^2 m^2$}, \quad
\footnotesize{$f_{(3a)1}=x_2-y_2,\quad f_{(3a)2}=y_1$}\\ \hline
$ 4a $&$\frac{i}{3} $&$
\hat a_{(4a)}=\frac{s^2\,t\,\alpha_s(-x_2 y_2 t)\,\alpha_2\,
\kappa\,(\alpha_1-y_1)\,(y_1-x_1)}{ m x_1 y_1 m_V^2}
[2 y_2 \alpha_1+4 x_2 \alpha_2$ \\[1.0ex]
&&$
+\kappa (3 y_1 \alpha_1-y_1-6 \alpha_1^2+8 \alpha_1+5 x_1 \alpha_1-4-5 x_1)]
F_V^{(4)}(-x_2 y_2 t)$ \\ [1.0ex]\cline{3-3}
&&
\footnotesize{$d_{(4a)1}=(\alpha_1-x_1) (\alpha_1-y_1) t+(x_1-y_1)^2
m^2$}\\
&&
\footnotesize{$d_{(4a)2}=-(\alpha_1-y_1) \alpha_2 t+y_2^2 m^2$}, \quad
\footnotesize{$f_{(4a)1}=x_1-y_1, \quad  f_{(4a)2}=y_2$}\\
\hline
$ 5a  $&$\frac{8}{27} $&$
\hat a_{(5a)}=\frac{2\,s^2\,t^2\,\alpha_s(-x_2 y_2 t)\,\alpha_1\,\alpha_2\,
\kappa\,(y_2-\alpha_1)^2 }{m m_V^2}
[y_2 \alpha_2+2 x_2 \alpha_1-\kappa (\alpha_1+y_2 \alpha_2-3 x_2 \alpha_1+4 \alpha_1^2)]
F_V^{(4)}(-x_2 y_2 t)$ \\[1.0ex] \cline{3-3}
&&
\footnotesize{$d_{(5a)1}=(\alpha_1-y_2) \alpha_1 t+y_2^2 m^2$},\quad
\footnotesize{$d_{(5a)3}=(\alpha_1-x_2) (\alpha_1-y_2) t+(x_2-y_2)^2 m^2$}\\
&&
\footnotesize{$d_{(5a)2}=(y_2-\alpha_1) \alpha_2 t+y_1^2 m^2$}, \quad
\footnotesize{$f_{(5a)1}=-y_2, \quad  f_{(5a)2}=y_1, \quad f_{(5a)3}=x_2-y_2$}\\
\hline
\end{tabular}
\caption[]{Color factors and of the functions $d_{ij}$,
$f_{ij}$ and $\hat a_i$ at $\beta_1=\alpha_1$ for sample graphs\\
(for definitions see text). The contribution from graphs 4a and 5a is actually
given\\ for subgraph 6a.}
\label{tab}
\end{table}
\renewcommand{\arraystretch}{1.0}
There is a group of graphs in which the large variable $s$ appears in
two propagators denominators ($i=2c,3a,3b,4a,4b$):
\begin{equation}
\label{int_2}
\hat A_i=\hat a_i(\alpha_1,\beta_1) \;
        \frac{1}{s (\alpha_1-\beta_1) f_{i1}+d_{i1}+{\rm i}\, \epsilon_1}
\cdot \frac{1}{s (\alpha_1-\beta_1) f_{i2}+d_{i2}+{\rm i}\, \epsilon_2},
\end{equation}
where $f_{ij}$ and $d_{ij}$ are functions of the momentum fractions
$\alpha_1, \beta_1, x_1, y_1$. Moreover, the $d_{ij}$ depend on $t$ and
$m^2$ too. Obviously, these terms in the $d_{ij}$ have to be kept
now. Otherwise the integrals in (\ref{ampl}) would not exist.
$\hat A_i$ can easily be integrated over
$\beta_1$ by using partial fractioning and the standard formula
\begin{equation}
\label{pvi}
       \frac{1}{z +{\rm i}\epsilon} = {\cal P} \frac{1}{z} -{\rm i} \pi
                                                \delta(z)
\end{equation}
where ${\cal P}$ denotes the principal value integral. In the
kinematical region of interest, namely $m^2, |t| <<s$, the principal
value part can be shown to be suppressed by $1/s$ as compared to the
$\delta$ function part. The $\delta$ function provides the condition
$\beta_1=\alpha_1 +{\cal O}(1/s)$ in this case. Hence, to leading
order in $s$, we approximate (\ref{int_2}) by
\begin{eqnarray}
\label{intm}
 \hat A_i
&\simeq& - \frac{{\rm i} \pi}{s}\, \hat
a_i(\alpha_1,\alpha_1) \, \delta{(\beta_1 - \alpha_1)} \nn\\[0.5ex]
      &&   \left[\frac{{\rm
       signum}(f_{i1})}{d_{i2} f_{i1} -d_{i1} f_{i2}
                        +{\rm i}\, \epsilon\,  {\rm signum}(f_{i1})}
             - \frac{{\rm signum}(f_{i2})}{d_{i2} f_{i1}-d_{i1} f_{i2}
                     - {\rm i}\, \epsilon\, {\rm signum}(f_{i2})} \right].
\end{eqnarray}
Representative examples of the functions $d_{ij}$ and $f_{ij}$ as well as of
the $\hat a_i$ are quoted in the table.

The other integrations appearing in (\ref{ampl}) have to be done
numerically using (\ref{pvi}) again. Since in general
${\rm signum}(f_{i1})$ is not equal to
${\rm signum}(f_{i2})$ the $\hat A_i$ have both real and imaginary
parts. An exception is the graph 2c where
$f_{(2c)1}=x_1$ and $f_{(2c)2}=y_1$. In this case the two principal
value integrals cancel and the leading contribution to
$\hat A_{2c}$ therefore simplifies to
\begin{equation}
\hat A_{(2c)} \simeq -\frac{2 \pi^2}{s}\, \hat a_{(2c)}(\alpha_1,\alpha_1)
         \,\delta(\alpha_1 -\beta_1)\;\delta(d_{(2c)2} f_{(2c)1} -d_{(2c)1}
         f_{(2c)2}).
\end{equation}
With the help of this new $\delta$ function a second integration in
(\ref{ampl}) can be immediately carried out.

The graphs 5a and 5b, comprising 4-point diquark vertex functions,
have $s$ in three propagators.  The contribution of these graphs
can be written in the form
\begin{equation}
\label{m5}
\hat A_i = \hat a_i(\alpha_1,\beta_1)
\prod_{j=1}^{3}\frac{1}{s (\alpha_1-\beta_1) f_{ij}+ d_{ij}+{\rm i}\, \epsilon_j}
\end{equation}
As an example we quote the functions $\hat a_{5a}$  for the graph 5a together
with the $d_{(5a)j}$ and $f_{(5a)j}$ in the table.
To leading order in $s$ these contributions are also dominated
by the imaginary parts of the propagator poles at $-d_{ij}/(s
f_{ij})$. Up to corrections of order $1/s$ this again implies
$\beta_1=\alpha_1$. Thus, we find for $i=5a,5b$
\begin{eqnarray}
\label{m3}
  \hat A_i &\simeq& - \frac{{\rm i}\pi}{s}\,\hat a_i(\alpha_1,\alpha_1)
        \, \delta(\alpha_1 -\beta_1) \nn\\[0.5ex]
        &&   \left [ \frac{{\rm signum}(f_{i1})}{d_{i2} f_{i1} -d_{i1}
         f_{i2} + {\rm i}\,{\rm signum}(f_{i1})\,\epsilon_2}\;
          \frac{1}{d_{i3} f_{i1} -d_{i1} f_{i3} +
                    {\rm i}\,{\rm signum}(f_{i1})\,\epsilon_3}\right. \nn\\
        &&      \left.\phantom{\frac{1}{1}} +\, (1,2,3)\,{\rm cyclic} \right ]
\end{eqnarray}
How to proceed from here should be obvious.

Finally let us discuss the graph 3c. A pole only appears in the
$s$-channel propagator and $\hat A_{(3c)}$ is of the form
\begin{equation}
\label{mcb}
\hat A_{(3c)}= \hat a_{(3c)}
\frac{1}{s (\alpha_1-\beta_1)(y_1-x_1) + d_{(3c)}+{\rm i}\, \epsilon}.
\end{equation}
It can be shown that  the leading log contribution from this graph
to the integral over $y_1, \beta_1$ in (\ref{ampl})  is dominated
by the region near $\alpha_1=\beta_1$ and $y_1=x_1$:
\begin{equation}
\label{ln_term}
\int_{0}^{1}d y_1 \int_{0}^{1} d\beta_1
         \frac{F(s,t,\beta_1,y_,...)}
         {s (\alpha_1-\beta_1)(y_1-x_1) + d_{(3c)}+{\rm i}\, \epsilon} \sim
         F(s,t,\beta_1=\alpha_1,y_1=x_1...)I(s),
\end{equation}
where F absorbs all terms appearing in (\ref{ampl}) including
$\hat a_{(3c)}$ and
\begin{equation}
\label{lint}
I(s)=\int_{0}^{1}d y_1 \int_{0}^{1} d\beta_1
\frac{1}{s (\alpha_1-\beta_1)(y_1-x_1) + d_{(3c)}+{\rm i}\, \epsilon}.
\end{equation}
Approximately this integral is given by
\begin{eqnarray}
I(s)\sim \int_{-1/2}^{1/2}d u \int_{-1/2}^{1/2} dv
\frac{1}{s u v + d_{(3c)}+{\rm i}\, \epsilon}+ {\cal O }( 1/s )= \\ \nonumber
 \frac{2}{s} \left[\mbox{dilog}\left(\frac{-s}{4 d_{(3c)}}\right)-
 \mbox{dilog}\left(\frac{s}{4 d_{(3c)}}\right)\right]
\sim -\frac{2 {\rm i} \pi}{s} \ln{s}.
\end{eqnarray}
Note, that $\hat a_{(3c)} \propto s^2$ as the contributions from the
other graphs (see the table).
Thus, the dominant contribution from graph 3c is
\begin{equation}
 (F_{+-})^{LL}_{(3c)} \propto {\rm i}\, s \, \ln{(s)}\,f(t).
\end{equation}
 We calculate numerically in (\ref{ln_term}) not only the leading $s
\ln{s}$ term but also the non-logarithmic contribution which
behave like $s$.

\section{Numerical results for spin-dependent $pp$ scattering}
In our numerical studies of proton-proton scattering
we use the following form of the scalar and
vector diquark DA
\begin{equation}
\label{a10}
\begin{array}{l}
\varphi_S(x_1)=N_S\, x_1 x_2^3\exp{\left[-b^2
(m^2_q/x_1+m^2_S/x_2)\right]}\\
\varphi_V(x_1)=N_V\, x_1 x_2^3(1+5.8\,x_1-12.5\,x_1^2)
\exp{\left[-b^2 (m^2_q/x_1+m^2_V/x_2)\right]}
\end{array}
\end{equation}
and the set of parameters
\begin{equation}
\label{c1}
\begin{array}{cccc}
 f_S= 73.85\,\mbox{MeV},& Q_S^2=3.22 \,\mbox{GeV}^2, & a_S=0.15,
 &   \\
 f_V=127.7\,\mbox{MeV},& Q^2_V=1.50\,\mbox{GeV}^2, &
 a_V=0.05,&\kappa=1.39\,
\end{array}
\end{equation}
as proposed in \cite{diquarks,Jak:93b}. The values of the masses
in the exponentials are taken as $330\,\mbox{MeV}$ (for the quarks)
and $580\,\mbox{MeV}$ (for the diquarks). The transverse size parameter
$b$ is taken to be $0.498\,\mbox{GeV}^{-1}$. The normalization constants $N_S$
and $N_V$ have the values 25.97 and 22.29, respectively.
As we mentioned in the preceeding section the $\beta_1$ integration
is trivial. The other three integrations over the hard amplitude and
the proton DAs are carried out numerically. Since we neglect $1/s$
corrections throughout we find an energy independent ratio of the
helicity-flip and non-flip amplitudes.

Let us discuss the role of the contributions from the individual
graphs briefly. The contributions from the graphs 2a and 2b  to $F_{+-}$ are
purely imaginary. Thus, although these contributions lead to helicity
flips they do not produce a phase difference between the $F_{+-}$ and $F_{++}$
and, hence, do not contribute to the single spin asymmetry.
The graph 2c yields a real contribution that is quite small, about
a few percent of Im $F_{+-}$ at $|t| \le 10 \,\mbox{GeV}^2$.
The contributions to the real part of $F_{+-}$
provided by the graphs 3a and 3b though substantial are compensated
by the contribution from graph 3c to a large extent.
The contributions of the graphs 4a, 4b, 5a and 5b
to the real part of $F_{+-}$ are very small as the numerical evaluation
reveals. Their imaginary parts, however, are not small as is that from
graph 3c. These imaginary contributions play an important role for the
double spin asymmetry parameter $A_{NN}$.

The results of our calculations for the helicity flips amplitude
$F_{+-}$ are shown in Fig.\ 7 for $s=100\,\mbox{GeV}^2$. As can
be seen from that figure the imaginary part of $F_{+-}$ is much
larger than its real part. The real part of $F_{+-}$ changes
sign at $|t| \sim 3.5 \,\mbox{GeV}^2$. The absolute value of the
ratio of helicity-flip and non-flip amplitudes is fairly large
$|F_{+-}|/|F_{+-}| \sim 0.2-0.3$ at $|t| \ge 3 \,\mbox{GeV}^2$
indicating the substantial amount of helicity flips generated
through the vector diquarks in our model.

The interference of the real part of $F_{+-}$ with the purely
imaginary ansatz for the amplitude $F_{++}$ yields the
single-spin asymmetry $A_N$ (\ref{an}). Our prediction for $A_N$
at $s=100\,\mbox{GeV}^2$ and for  $|t|\geq 3  \,\mbox{GeV}^2$ is
shown in Fig.\ 8 and compared to the only available experimantal
data in that region (at $s = 370 \,\mbox{GeV}^2$) \cite{fnalp}.
The quality of the present data is poor and prevents any severe
test of our predictions. The predicted asymmetry amounts to
about 20--30\% for $|t| > 6 \,\mbox{GeV}^2$; it is of the same
order of magnitude as has been observed in the low-energy  BNL
experiment \cite{krish}. The decrease of the asymmetry at
smaller momentum transfer is connected with  the smallness of
${\rm Re} F_{+-}$ near $|t| =3 \,\mbox{GeV}^2$.

The predictions for the double spin asymmetry $A_{NN}$ are shown in
Fig.\ 9. $A_{NN}$ turns out to be of the order of $10-20\%$.
Our results for the spin asymmetries are rather close to those
obtained in \cite{gol-91,gol-95} although the latter are valid in the
momentum transfer region $2\,\mbox{GeV}^2 <|t| < 4\,\mbox{GeV}^2$.
The spin observables obtained within the model
are essentially independent on the parameterizations
(\ref{mpe},\ref{pinch}) used for the non-flip amplitude $F_{++}$.
\section{Summary}
On the basis of the diquark model we have calculated  spin
effects in high-energy proton-proton scattering at moderately
large momentum transfer.  The two-gluon graphs for the
colour--singlet $t$-channel exchange have been considered for
the helicity flip amplitude while for the helicity non-flip
amplitude a phenomenological  parameterization is used. It
describes qualitatively the differential cross section of the
elastic $pp$ scattering. The $F_{+-}$ amplitude is calculated
under the assumption that the t-channel gluons couple to one
constituent, quark or diquark, each in the helicity non-flip
vertex. In the helicity flip amplitude we include the
perturbative $\alpha_s$ correction. Hence, we consider minimally
connected graphs which allow to keep all constituents collinear.
In our model the helicity flips are generated by vector
diquarks. It turns out that the flip amplitude  $F_{+-}$ is of
substantial magnitude and not in phase with the non-flip
contribution.

Our model, therefore, provides a single-spin asymmetry
that is rather large for momentum transfer $|t| \ge 3\,
\mbox{GeV}^2$.
The double spin transverse asymmetry in this kinematical region
are rather large in our model.
The important feature of the spin effects obtained in our model is
their weak energy dependence. On the other hand, they decrease with
increasing  momentum transfer. Our results are valid at large $s$
and moderately large momentum transfer ($>  $few GeV$^2$). This
kinematical region can be investigated for instance in the proposed
HERA-$\vec N$ experiment \cite{anselm}.

Finally we want to stress that our predictions for $A_N$ should not be taken
literally since phase differences are hard to calculate, they depend
on many subtle details which are not well under control in a model.
The diquark model on which our model is based
was designed for a different kinematical region.
In so far, a failure of our prediction for $A_N$ would not necessarily
imply a failure of the diquark model in general but would rather indicate
that the phase differences are not well under control and/or that the
diquark model is applied beyond its range of applicability.

\section{Acknowledgments}
We would like to thank J.Bolz, A.Efremov, R.Jakob, O.Nachtmann
for fruitful discussions. S.V.G. would like to thank the
Fachbereich Physik, Universit\"at Wuppertal, for the warm
hospitality in Wuppertal. This work was supported in part by the
Heisenberg-Landau Grant.


\phantom{.}
\centerline{\bf Figures}
\noindent
FIG.1.\ Structure of the spin-non-flip proton vertex.\\
FIG.2.\ Feynman graphs containing the 3-point diquark function
(without 3-gluon coupling).\\
FIG.3.\ Feynman graphs containing the 3-point diquark function
(with 3-gluon coupling).\\
FIG.4.\ Feynman graphs containing the 4-point diquark function
(with 3-gluon coupling).\\
FIG.5.\ Feynman graphs containing the 4-point diquark function
(without 3-gluon coupling).\\
FIG.6.~Structure of the 4-point diquark function.\\
FIG.7.~t-dependence of the $F_{+-}$ amplitude at
$s=100 \mbox{GeV}^2$, solid line-imaginary
part; dot-dashed line-real part.\\
FIG.8.~Model predictions for single-spin asymmetry at
$s=100 \mbox{GeV}^2$ (solid line: for
the MPE model (\ref{mpe}); dashed line: for the LP model
(\ref{pinch})).\\
FIG.9.~Model predictions for double-spin asymmetry at
$s=100 \mbox{GeV}^2$ (solid line: for
the MPE model (\ref{mpe}); dashed line: for LP model (\ref{pinch})).

\newpage
\phantom{.}
\vspace*{-1.9cm}
\hspace*{.1cm}
\epsfxsize=15.5cm
\epsfbox{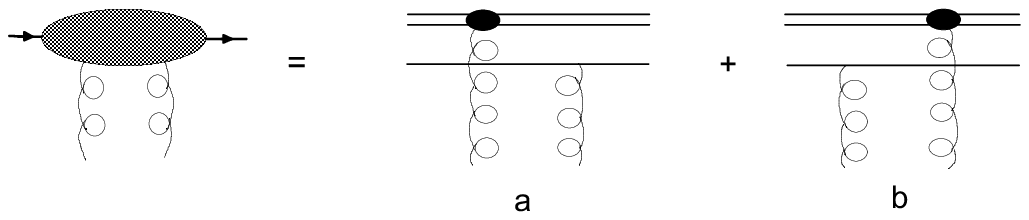}
  \vspace*{-19.cm}
\begin{center}
Fig.1
\end{center}

\nopagebreak[4]
  \vspace*{-1.cm}
\hspace*{.4cm}
\epsfxsize=15.4cm
\epsfbox{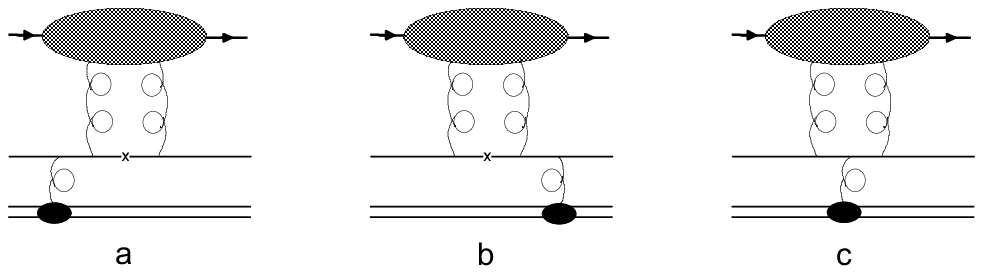}
  \vspace*{-19.cm}
\begin{center}
Fig.2
\end{center}

 \vspace*{-1.0cm}
\epsfxsize=15.5cm
  \hspace*{-.6cm}
{\epsfbox{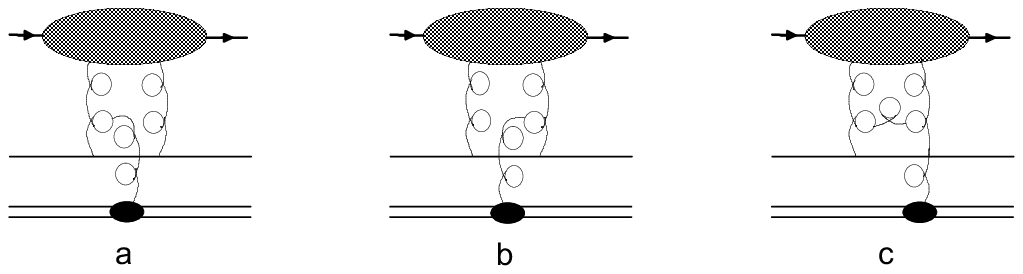}}
  \vspace*{-18.5cm}
\begin{center}
Fig.3
\end{center}

  \vspace*{-1.0cm}
\epsfxsize=16cm
  \hspace*{-1.0cm}
 {\epsfbox{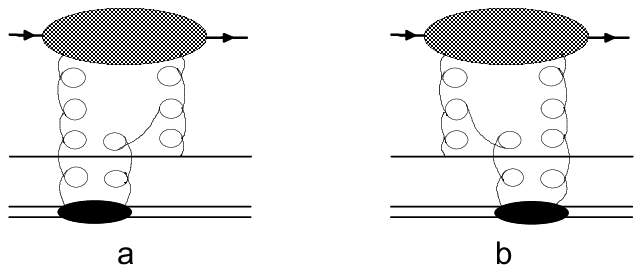}}
 \vspace*{-19.0cm}
\begin{center}
Fig.4
\end{center}

 \vspace*{-1.0cm}
 \hspace*{.3cm}
\epsfxsize=15.5cm
\epsfbox{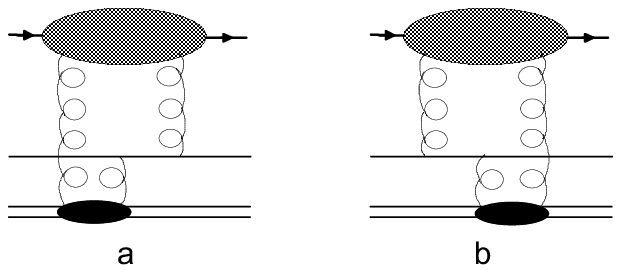}
  \vspace*{-19.2cm}
\begin{center}
Fig.5
\end{center}

\vspace*{-1.1cm}
\epsfxsize=16cm
\centerline{\epsfbox{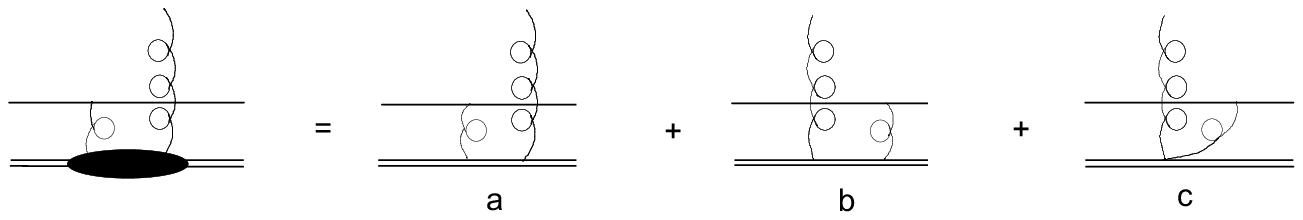}}
  \vspace*{-19.2cm}
\begin{center}
Fig.6
\end{center}

\newpage

  \vspace*{-.5cm}
       \hspace*{.9cm}
\epsfxsize=13cm
{\epsfbox{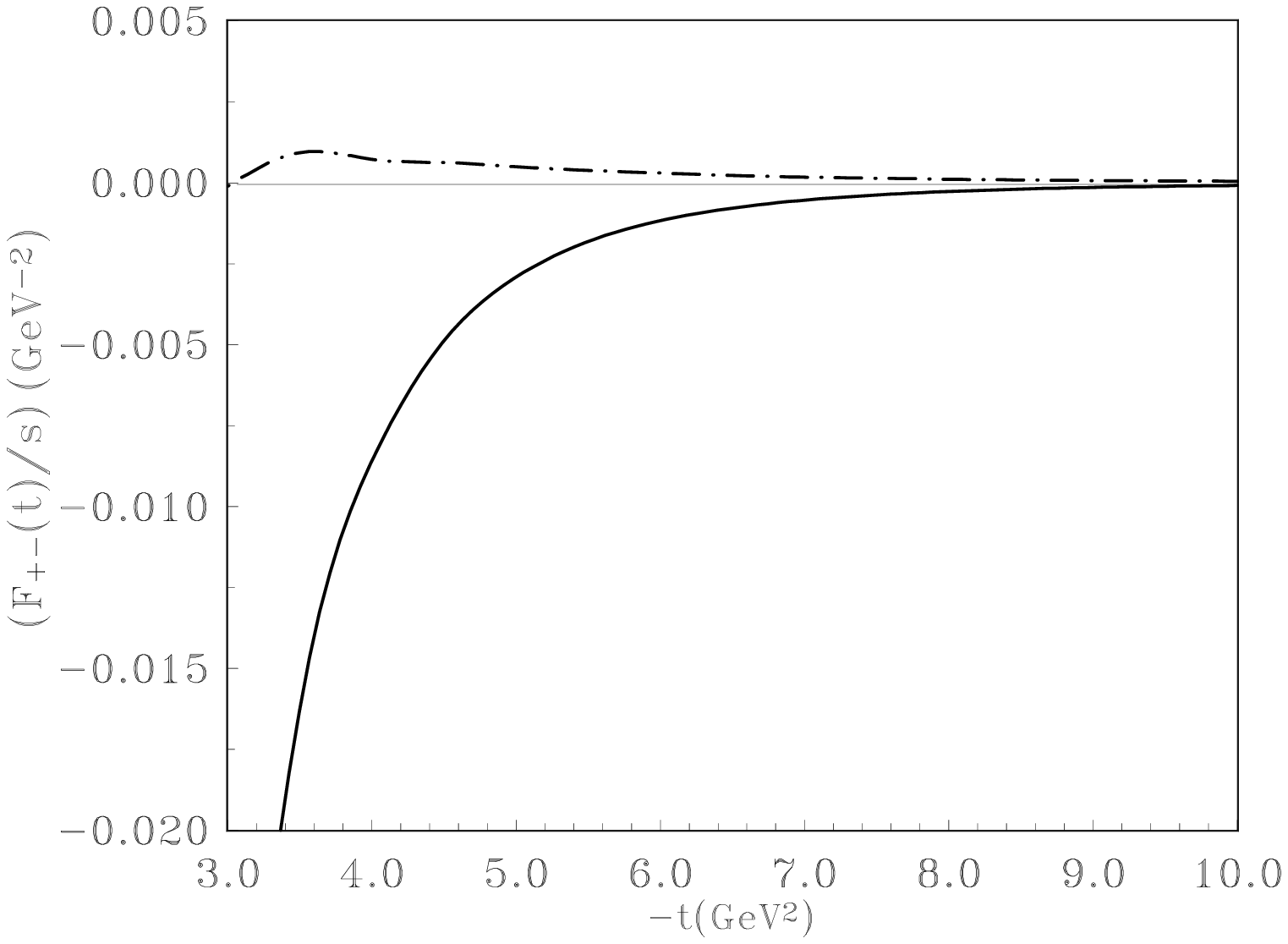}}
  \vspace*{.3cm}
\begin{center}
Fig.7
\end{center}

  \vspace*{-.1cm}
\epsfxsize=13cm
\centerline{\epsfbox{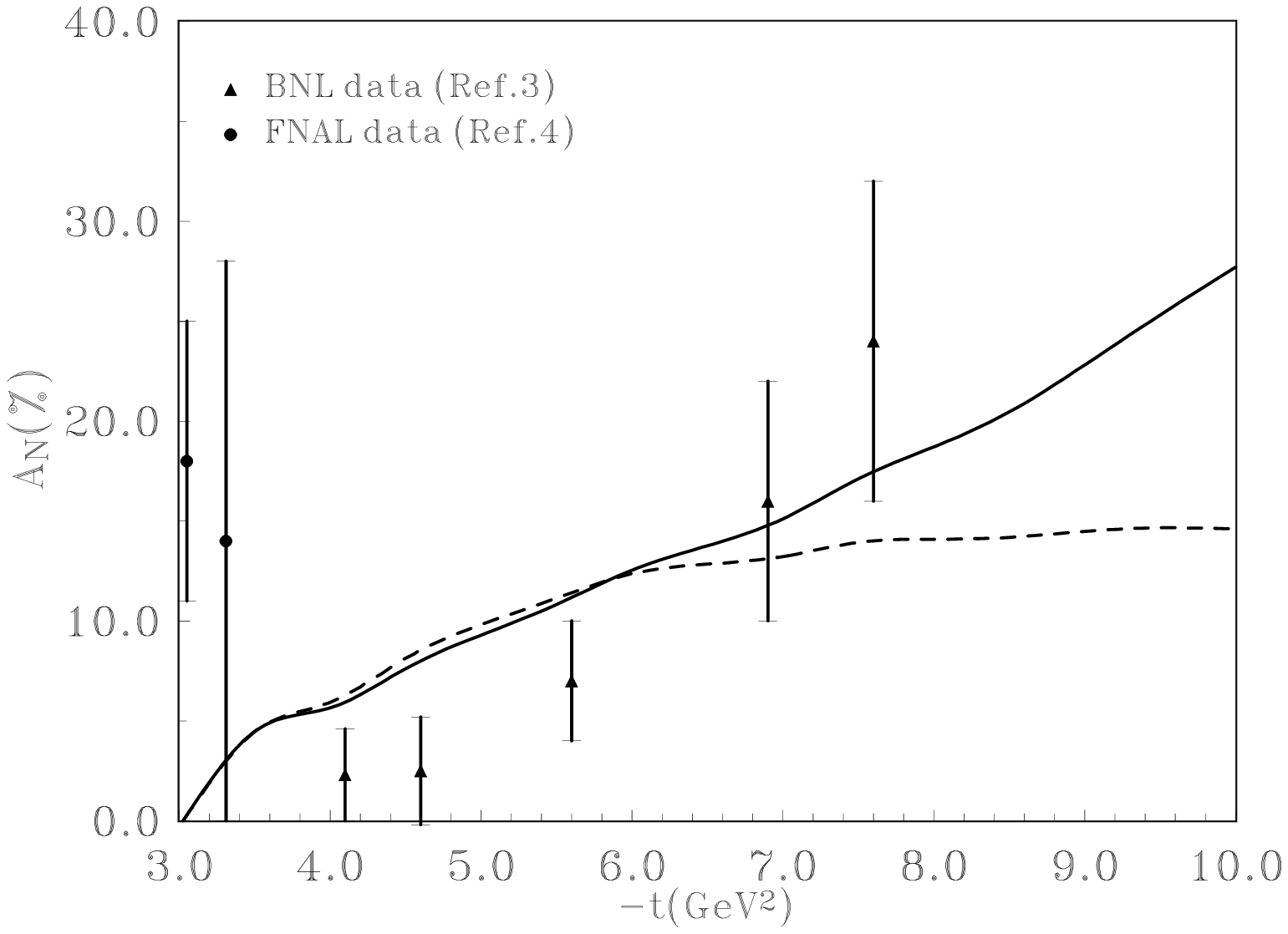}}
  \vspace*{.2cm}
\begin{center}
Fig.8
\end{center}

\newpage
  \vspace*{-.5cm}
  \hspace*{.9cm}
\epsfxsize=13cm
{\epsfbox{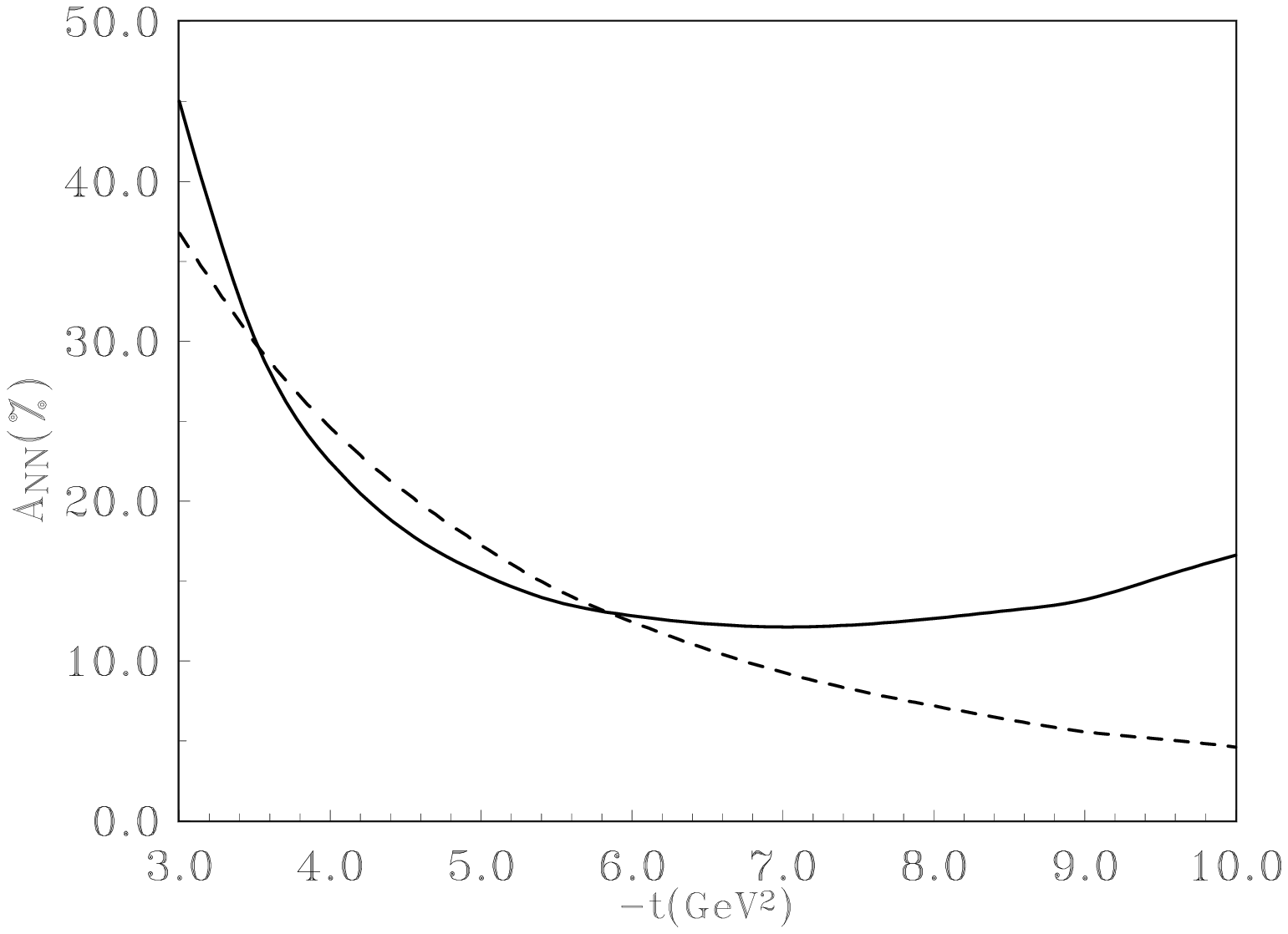}}
  \vspace*{.3cm}
\begin{center}
Fig.9
\end{center}

\end{document}